\documentclass{sigplanconf}

\usepackage{epsfig}
\usepackage{amsmath}
\usepackage{extarrows}

\begin{document}

\setlength{\pdfpageheight}{\paperheight}
\setlength{\pdfpagewidth}{\paperwidth}


\toappear{%
\noindent Permission to make digital or hard copies of all or part of this 
work for personal or classroom use is granted without fee provided that copies are
not made or distributed for profit or commercial advantage and that copies bear this
notice and the full citation on the first page. \par 
\vspace{2pt} 
\noindent \textsl{TaPP 2016}, \quad June 8--9, 2016, Washington, DC. \par 
\noindent Copyright remains with the owner/author(s).}


\titlebanner{banner above paper title}        
\preprintfooter{short description of paper}   

\title{Automatic vs Manual Provenance Abstractions: Mind the Gap}

\authorinfo{Pinar Alper}
           {University of Manchester}
           {alperp@cs.manchester.ac.uk}

\authorinfo{Khalid Belhajjame}
           {Universit\'{e} Paris Dauphine}
           {Khalid.Belhajjame@dauphine.fr}

\authorinfo{Carole A. Goble}
           {University of Manchester}
           {carole.goble@manchester.ac.uk}

\maketitle

\begin{abstract}
In recent years the need to simplify or to hide sensitive information in provenance has given way to research on provenance abstraction. In the context of scientific workflows, existing research provides techniques to semi-automatically create abstractions of a given workflow description, which is in turn used as filters over the workflow's provenance traces.  An alternative approach that is commonly adopted by scientists is to build workflows with abstractions embedded into the workflow's design, such as using  sub-workflows. This paper reports on the comparison of manual versus semi-automated  approaches in a context where result abstractions are used to filter report-worthy results of  computational scientific analyses. Specifically; we take a real-world workflow containing user-created design abstractions and compare these with  abstractions created by  ZOOM*UserViews and Workflow Summaries systems.  Our comparison shows that semi-automatic  and manual approaches largely overlap from a process perspective, meanwhile, there is a dramatic mismatch in terms of data artefacts retained in an abstracted account of derivation. We discuss reasons and suggest future research directions.
\end{abstract}








\section{Introduction}

Provenance brings transparency into  past processes, which is crucial for their audit/verification  or for establishing  the quality and trustworthiness of  their results.  Transparency, on the other hand, is a double-edged sword as provenance can at times be considered  too revealing  or too detailed description of a process.  Side effects of transparency is counterbalanced by Provenance Abstraction, for which there are two major motivations. First is privacy and security, where  a system's execution traces may need to be redacted to protect data confidentiality, or hide operational vulnerabilities \cite{perera-cheney}. The second driver is simplicity \cite{Biton2008}. Provenance records, especially those automatically collected from monitored execution of systems -be them  databases, workflow engines or file systems- are known to be voluminous and complex \cite{Chapman:2008:EPS:1376616.1376715}. \textit{In the context of this paper we focus on workflow provenance and analyse abstractions generated with the goal of simplicity}.

With complexity we refer to the structural complexity of provenance graphs documenting causal relations among computational processes, sub-processes and the numerous data generated. In the context of scientific workflows, complexity is rooted in the complexity of workflow descriptions.  Empirical analyses show that workflows can contain up to 50+ data processing steps \cite{DBLP:journals/fgcs/GarijoABCGG14}. Another empirical reality is that the majority of steps (up to 70\%) are dedicated to data-adaptation, whereas a minority performs scientifically significant processing.  

When complex workflows are executed they generate a large number of intermediary and final outputs, which are interlinked with deep lineage paths, which can be a barrier for the exploitation of provenance  by human-users.  One example is workflow debugging, where the user has to navigate through lineage for  data validation \cite{Biton2008}. Here shortening of lineage can speed up the isolation of errors.  Another example is the reporting of workflow-based computational experiments  \cite{belhajjame-mina}. Abstraction is desired here as adapters can obfuscate the scientific intent of the analysis. Reporting involves the creation of bundles (zip files) of resources associated with an experiment including workflows, their input/output data and provenance metadata.  Observation on existing bundles show that scientists typically share large provenance graphs in their entirety as proof of conduct of their experiment; meanwhile, they  also share  an abstracted view of the activities that make up the analytical process and a selected subset of  data items and their dependencies.  \textit{In this paper we analyse suitability of abstractions for experiment reporting.}

Broadly, there can be two strategies for abstraction: 
\begin{enumerate}
\item Preempting complexity by manually encoding abstractions into the design of the computational instrument used for data processing. For workflows,  this is achieved using design constructs such as  sub-workflows. Design abstractions, as the name implies, are embedded in design and, therefore they are static for a particular workflow (a particular version developed by a particular user). Once design abstractions are created they can be used several times for all executions of that workflow. Scientists typically create design abstractions to later exploit them in reporting. 
\item Devising abstractions post-hoc, either over completed workflow designs or provenance recorded from executions. The state of the art research on (semi)automated provenance abstraction fall in this category. Unlike design abstractions, post hoc abstractions can be dynamic, meaning different abstractions can be created over the same workflow  or execution trace.  
\end{enumerate}

In this paper we compare the above two strategies. Workflows publicly shared in repositories, such as myExperiment \cite{myExp2008} provide examples of the first strategy. For the second  i.e. the semi-automated category we use two abstraction  systems described in literature  namely \textit{ZOOM*UserViews} and \textit{Workflow Summaries}.  We have selected these as they are representative techniques for abstracting workflow provenance with the goal of simplification. ZOOM system exploits user-supplied abstraction hints identifying significant activities in a workflow. The Workflow Summaries system exploits  annotations, which may have been provided for purposes other than abstraction, to identify significant elements of the workflow. 

We begin by illustrating design abstractions (first category) in Section \ref{sec:des-abs}.  Following that in Section \ref{sec:auto-abs} we dissect  provenance abstraction to its basic components and describe the two systems (second category)  used in our comparison. We present our methodology in applying two abstraction systems over the same workflow description and outline our measure in comparing abstractions and present results in Section \ref{sec:results}. We discuss the factors that shape  abstractions and  outline future research directions in Section \ref{sec:discuss}. We conclude  in Section \ref{sec:conclusion}.

\section{Workflow Design Abstractions}
\label{sec:des-abs}

Scientists  create  design abstractions in two ways;  one is by using  \textit{sub-workflows}, and the other is by \textit{bookmarking output ports of activities}. 

\begin{figure}
\begin{center}
\includegraphics[scale=0.35]{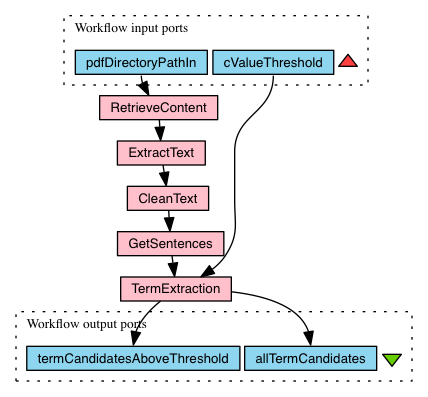}
\end{center}
\caption{A Text Mining Workflow Comprised of Sub-Workflows.}
\label{fig:wf-nested}
\end{figure}

Managing complexity of a design artefact with hierarchies/layers is  an established technique in the fields of software engineering or business process modelling \cite{bpm-layers}.  Sub-workflows allow for layered designs, and are a best-practice when building large workflows \cite{HettneWBGMDRVGR12}.  Figure \ref{fig:wf-nested}  displays a text mining workflow obtained from myExperiment\footnote{http://www.myexperiment.org/workflows/1061.html}. The workflow is comprised of 5 sub-workflows  dedicated to \begin{scriptsize}(1)\end{scriptsize} retrieval of contents from a list of file names, \begin{scriptsize}(2)\end{scriptsize}  extraction of text from content, \begin{scriptsize}(3)\end{scriptsize}  cleaning of text,  \begin{scriptsize}(4)\end{scriptsize}  extraction of sentences  and \begin{scriptsize}(5)\end{scriptsize}  the detection of terms within this corpus. When the sub-workflows are expanded (grey-shaded boxes in Figures \ref{fig:collapse-w-example},\ref{fig:zoom-example}, and \ref{fig:collapse-n-example} in the Appendix) we can observe 19  activities in total. The text-mining activities are realised by calls to a web service, in addition there are several adapters realised by local scripting and XML processing tools.  As a result  of the use of sub-workflows the design of the process as observed at the top layer  has been significantly simplified containing less number of  activities, ports, and dataflow links among those ports. Consequently when executed the top layer of design presents a more compact data lineage. The activity groupings that  make up sub-workflows can be determined by various criteria; the most common, which is also  illustrated in our example, is Functional Modularity.  We can observe from the sub-workflows in Figures \ref{fig:collapse-w-example},\ref{fig:zoom-example}, and \ref{fig:collapse-n-example} that major analytical steps of the workflow are grouped with related adapters responsible for the preparation of the inputs  parameters  for an analysis and the post-processing of outputs. 

The second type of design abstraction is promoting activity output ports holding (intermediary) results to become workflow outputs. We refer to this  pattern as \textit{lineage bookmarks}, others have named it ``trace links'' \cite{cohen-distill}.  The sub-workflow named $TermExtraction$ in Figures \ref{fig:collapse-w-example},\ref{fig:zoom-example}, and \ref{fig:collapse-n-example} illustrates this pattern. Here the output of activity $Concatenate\_two\_strings\_2$ is an intermediate value used by a downstream activity named $jamesXPath$. The scientist has chosen to promote this intermediate port to become an output of $TermExtraction$, named $xpathOutput$. This way data, which would otherwise be hidden in a sub-workflow, has become visible at the top layer of design.

There exist significant past research and tooling on aiding scientists to discover analysis or adapter components from respective catalogues and compose them as workflows  \cite{sadi-semantic}.  On the other hand to our knowledge there exist no feature in state of the art workflow systems that aid in modularisation or refactoring of large workflows. Consequently, creation of design abstractions is a common yet manual process.

\section{Abstraction as a Process}
\label{sec:auto-abs}

From a high level viewpoint abstraction systems take as input a \textbf{provenance graph} and an \textbf{abstraction policy}  and produce an \textbf{abstracted provenance graph}. The policy identifies graph parts to be retained or abstracted-away. It may additionally specify how abstraction should occur, i.e. how input graph should be manipulated.  A large majority of abstraction systems operate over retrospective provenance, for systems specialised on workflow provenance, however, abstraction typically occurs over prospective provenance, i.e. workflow descriptions. 

Abstraction process may also commit to a number of  \textbf{integrity policies}, which are often determined by the end-purpose of abstractions. Integrity policies shape how informative,  valid and well-formed result abstraction will be.  Herein we informally outline common integrity policies, and use workflow design abstractions to illustrate each (see  Table \ref{table:abst-approach} for a summary): \begin{itemize}
\item Abstraction  preserves dependency \textbf{soundness} when its result contains no false data dependencies, which are those unfounded (not backed) by a dependency in the original provenance graph. Abstraction based on the grouping of nodes, as in the creation of sub-workflows,  does not preserve soundness. Consider the sub-workflow  named $TermExtraction$ given  in Figures \ref{fig:collapse-w-example},\ref{fig:zoom-example}, and \ref{fig:collapse-n-example}. From the expanded view we can observe that the input  of  $TermExtraction$ named $sentencesList$  does not contribute to the creation of output named $XPathOutput$. However as per abstraction, at the top layer  of workflow design $TermExtraction$  is a single activity hiding precise dependencies among its inputs and outputs and giving the impression that all of its inputs contribute to all its outputs.
\item Abstraction  preserves  \textbf{acyclicity} when abstraction actions do not introduce cycles of dependency relations among data nodes. Our example workflow given in Figures \ref{fig:collapse-w-example},\ref{fig:zoom-example}, and \ref{fig:collapse-n-example} contains no cyclic dataflows. Some workflow systems such as Kepler and Taverna (from which our example comes), support a special kind of cyclic dependency, called a feedback loop (where an activity that is ran by an initial seed generates output that is consequently used as the next input seed) to achieve iteration. Iterated activity invocations result in acyclic retrospective execution provenance graphs. Existing scientific workflow systems require that all inputs of an activity are available to initiate its execution. Therefore beyond the special case of feedback loop, workflow systems do not allow sub-workflows containing cyclic dependencies where a dataflow path that goes  out of a sub-workflow can be traced back into it.
\item Depending on the source system, from which provenance is collected, patterns may exist in provenance. Most noted  pattern in the context of workflow provenance is \textbf{bi-partiteness}, where the account  of data processing  is given in the form $\textit{data} \xLongrightarrow{\text{{generatedBy}}}  \textit{activity} \xLongrightarrow{\text{{used}}}  \textit{data}$.  In such traces lineage among data items is always contextualised by some (data processing) activity in-between.  Design abstractions created by the sub-workflow construct preserve bi-partiteness.
\item Abstraction preserves \textbf{validity} if its result is some valid provenance graph, as per existing  models of provenance e.g. OPM, PROV. Models bring restrictions on the kinds of nodes that can occur in a provenance graph and their allowed relations. Abstractions based on free-style node grouping, or graph manipulation,  particularly seen in security-driven scenarios, may break validity. One example is provenance redaction \cite{Cadenhead:2011:TPU:1998441.1998456},  where a group of activity, data and actor nodes may be replaced by an opaque censor node that is typeless (i.e.  does not correspond to any valid provenance element). Meanwhile, approaches in workflow provenance abstraction typically  preserve validity.
\item Abstraction preserves  dependency \textbf{completeness} if it preserves all depedency relations among data nodes that exist in both the original and abstracted graph.  Abstractions based on node grouping typically preserves completeness, whereas those based on unrestricted elimination of edges and nodes may not. Similar to validity, completeness is often compromised in security-driven abstraction \cite{Cadenhead:2011:TPU:1998441.1998456}, whereas it is preserved in workflow provenance abstraction. 
\end{itemize}

\begin{table}[ht] 
\begin{scriptsize}
\caption{Workflow Abstraction Approaches} 
\centering \begin{tabular}{p{1.7cm}|p{1.1cm}|p{1.1cm}|p{1.1cm}|p{1.4cm}}
\hline\hline \\
\textbf{Approach} & {ZOOM \par UserViews}& \multicolumn{2}{|c|}{\parbox[c]{2.2cm}{Workflow  Summaries}} & {Design \par Abstractions}\\ \hline	
\textbf{Method} & Composite & Collapse & Eliminate & Sub-Wf \\ \hline
\textbf{Soundness} & Y & N & Y & N \\ 
\textbf{Acyclicity} & N & Y & Y & Y \\ 
\textbf{Bipartiteness} & Y & Y & N & Y \\ 
\textbf{Validity} & Y & Y & Y & Y \\ 
\textbf{Completeness} & Y & Y & Y & Y \\ \hline\hline
\end{tabular} 
\label{table:abst-approach} 
\end{scriptsize}
\end{table}

We will now describe the two abstraction systems used. 

\textbf{ZOOM} \cite{Biton2008}  accepts as abstraction policy the list of activities  that a user deems significant within a workflow. This information is then used to group activities into composites, similar to sub-workflows, where each composite contains at most one significant activity.  The result  is called a User's View over the original workflow. Based on input policy different views over the same workflow can be created. In addition to the user-specified activities, as a built-in abstraction policy, ZOOM treats all workflow inputs and outputs also  as significant items.  The integrity policy built into the ZOOM system is designed to preserve soundness of dataflow relations among significant items when creating composites.  On the other hand, ZOOM permits creating cyclic abstractions.

\textbf{Workflow Summaries} \cite{summaryBigData13} provides two different abstraction methods. First is based on activity grouping using a method named Collapse. The second method is controlled activity Elimination, where the incoming and outgoing dataflow links of the eliminated activity are replaced by indirect links, thereby preserving data dependency completeness. This system exploits Motifs, which are semantic annotations of workflow activities designating their functionality (e.g. Filtering, Format Transformation etc).  The abstraction policy in this system is a list of motif-action pairs identifying which  adapters shall be abstracted away using which method. As the policy refers to activity functionalities rather than individual activities, it can be used across multiple workflows.  Depending on the abstraction method there can be differences in integrity guarantees. The Eliminate method cannot provide a bipartite account of derivation as it introduces indirect dataflow links. The Collapse method is based on grouping therefore it does not preserve soundness. On the other hand this system preserves acyclicity in both Eliminate and Collapse methods.

\section{Comparing Abstractions}
\label{sec:results}

In order to compare (semi)automatically generated abstractions with each other and against user-generated design abstractions, we performed a test run. To prepare the input to the abstraction process we flattened the workflow in Figure \ref{fig:wf-nested} by un-nesting sub-workflows.  We prepared policies for each system as follows:\begin{itemize}
\item For Workflow Summaries we annotated workflow activities with Motifs. We prepared three policies given  in Table \ref{table:abst-precision}, namely Collapse-All, Eliminate-All, and Collapse-Selected, which prescribe respectively  \begin{scriptsize}(1)\end{scriptsize}  the grouping of any kind of adapter with  its upstream, if not possible, downstream activity  \begin{scriptsize}(2)\end{scriptsize} the elimination of all adapters   \begin{scriptsize}(3)\end{scriptsize} the grouping of selected kinds of adapters (discussed later in this section) with upstream or downstream activities.
\item For  ZOOM  we designated analytical (non-adapter) activities as significant, these activities are denoted with black stars in Figures 2, 3 and 4. 
\end{itemize}
 
We assess semi-automatically generated abstractions by using the user-generated design abstractions as ground truth. One question that may arise is ``why would a user abstraction be representative of the ground truth?''. As identified earlier, experiment reporting is our  particular focus. Currently design abstractions are  the only abstraction mechanism utilised for reporting. Our analyses of repositories have shown that shared workflows either contain no design abstractions, or they are highly abstracted (with sub-workflows). In our analyses we did not encounter a case where the same workflow had two different design abstractions made by different users. Therefore we deem a particular user's  design abstraction  for a particular workflow may be used as ground truth for that workflow in the context of reporting. The text mining workflow used in this paper was designed by an experienced user who had  multiple submissions to myExperiment.

We use elements of the main data derivation path in Figure 1 for comparison (path from input $pdfDirectoryPathIn$ to output $termCandidatesAboveTreshold$). For each abstraction we look at process and data-wise overlaps with the ground truth as follows (also displayed in Table \ref{table:abst-precision}).  Process-wise, we measure   \textbf{activity}  precision  denoted  with A/B. Where B  represents the total number of activities in the abstraction, and A denotes activities in the abstraction that have a correspondent in the design abstraction in Figure \ref{fig:wf-nested}. Similarly  we measure \textbf{activity port} precision to understand data-wise overlaps. 

\begin{table}[ht] 
\begin{scriptsize}
\caption{Precision of  Abstractions} 
\centering \begin{tabular}{p{3.5cm}|p{0.9cm}|p{1.3cm}|p{1.2cm}}
\hline\hline \\
\textbf{PolicyName} & \textbf{Activity}& \textbf{Activity Ports}  & \textbf{Illustrated In}\\ \hline	
{Wf Summaries -Eliminate All} & 5/5   & 0/9  & - \\ \hline
{Wf Summaries -Collapse All} & 5/5  & 0/4 & Figure \ref{fig:collapse-w-example} \\ \hline
{Wf Summaries- Collapse Selected} &  5/9 & 4/8  & Figure \ref{fig:collapse-n-example} \\ \hline
{ZOOM} & 5/7  & 0/6 & Figure \ref{fig:zoom-example}  \\  \hline\hline
\end{tabular} 
\label{table:abst-precision} 
\end{scriptsize}
\end{table}

Abstraction by Elimination of adapters is equivalent to hopping over  a data derivation path visiting only the scientifically significant activities and their direct inputs/outputs. Process-wise such an account overlaps fully (5/5) with the design abstraction of  Figure \ref{fig:wf-nested} as the user has created one sub-workflow per (significant) analytical activity. Data-wise, however, this method has reduced abstraction power, as it can only reduce the number of ports on the path to 9. A dataflow link in a workflow corresponds to a single data artefact in execution provenance, fulfilling roles of output of one activity and the input of the other. When we jump over traces (via indirect dataflow links) the two ends of such links corresponds to distinct artefacts. As a result the derivation path is less compact data-wise. More importantly, the elimination method has zero data-wise overlap with design-abstractions (0/9), meaning that none of the ports retained in the abstraction are those visible at the top layer of the design abstraction in  Figure \ref{fig:wf-nested}. Given that design abstractions are typically used to assist reporting, we understand that a significant activity and its report worthy output are not necessarily co-located, instead they may be separated by multiple activities in a workflow. Consider the web-service based text-mining activities in our example workflow comprised of sub-workflows. Neither outputs of  these service-based activities are visible at the top level of workflow design. Instead these activities are grouped with extractor type adapters that strip results from their service specific XML packaging. Therefore it is the output of these activities that gets reported.

Next we look  at  abstractions  based on grouping, specifically the ZOOM system and the Collapse-All policy of Workflow Summaries (black overlays in Figures \ref{fig:zoom-example}  and \ref{fig:collapse-n-example}).  Process-wise abstractions are highly similar to the design abstraction (5/5 for Collapse-All, 5/7 for ZOOM). Both systems have created activity groups, each involving one  analytical activity defining the overall function of that group. The process conveyed through these groups overlaps with the process in the user design. On the other hand, ZOOM system creates 2 groups, not containing analytical (significant) activities, denoted with solid boxes in  Figure \ref{fig:zoom-example}. These are due to the soundness policy of  ZOOM, to retain the information that workflow inputs and outputs have dataflow links among them do not pass via the analytical activity $analyze$.  When we look at the boundaries of groups we can observe both policies have caused a sweeping of adapters to nearby analytical activities that make up the boundary of groups. While there is an overlap from a Process perspective this sweeping style grouping presents stark data-wise mismatch against design abstractions (0/4 for Collapse-All and 0/6 for ZOOM). Visually this can be observed in the difference between the boundaries of system generated groups (black overlays) versus sub-workflows in Figures \ref{fig:zoom-example} and  \ref{fig:collapse-n-example}.

The Workflow Summaries system allows for more specific policies, and therefore more control over the abstraction process. We used this capability to encode a widely-observed  Functional Modularity design criteria that scientists adopt during workflow development \cite{DBLP:journals/fgcs/GarijoABCGG14}.   Analytical activities  that are handled by web services  typically require well-formed input requests (e.g. XML messages). Certain adapters in workflows are dedicated to this Input Preparation. There are also Extractor type adapters that process outputs of web services. The Collapse-Selected policy prescribes that Input Preparation adapters be grouped with downstream activities, and the Extractors with upstream. The resulting abstraction is given in Figure \ref{fig:collapse-w-example}. As we target only specialised adapters  smaller groups are generated and in turn a longer path, comprised of 9 activities, remains.  As a result of this less aggressive more mindful abstraction  half of the ports retained in the abstraction overlap with the  design abstraction (4/8).

\section{Discussion}
\label{sec:discuss}

Our analysis shows that semi-automated abstraction systems are focused on the process perspective as their input policies are predicated solely on activity significance/insignificance. As a result  of this approach resulting abstractions provide a simplified account of the analytical process but fail in supporting that account with data that fits in with the story implied by the process.  Consider the Workflow Summaries' abstractions (black overlays in Figure \ref{fig:collapse-n-example}): process-wise the first group represents the retrieval of a PDF document from the file system. Meanwhile the output of this group is not just the file content, as one would expect, it is instead a particular encoding of content embedded in an XML message to be sent to an external web service for text extraction (next group).  This mismatch would render automatically generated abstractions of little use in reporting scenarios and therefore highlights that the data-perspective needs to be taken into account.

An abstraction system's integrity policies are often determined by the end-use of result abstractions. The Soundness policy of the ZOOM system is designed to assist debugging scenarios. In our example this policy resulted in adapter-only activity groups as  all workflow inputs/outputs  and their dataflow dependencies deemed significant. The majority of inputs in our example workflow are configuration parameters, or implementation settings not particularly important for reporting. Consequently, this built-in assumption of input/output significance and the soundness  policy  renders ZOOM abstractions less suited for reporting. A similar observation can be made on acyclicity. While ZOOM abstractions can be cyclic, such views correspond to an account that is not technically feasible with existing workflow systems. This calls for more flexible  abstraction systems with configurable integrity policies or ones that can generate multiple abstractions exploring the spectrum of policy combinations. The ProPub system \cite{Dey:2011:PTD:2032397.2032414} from secure provenance literature adopts a foundational approach along this line. However, this system requires abstraction policies to refer to individual nodes in a retrospective provenance graph, which limits its usability in the context of abstracting scientific workflows.

Given that scientists are held accountable for reported work products (data and workflows), abstraction particularly in the context of reporting can be considered as a process in which the user must be involved and has the final say on the abstraction created. We believe this situation calls for further research into rethinking abstraction as a pre-hoc process that supports workflow design. One possible direction of future work could be in exploiting abstractions in existing similar workflows \cite{Starlinger:2014:SSS:2732977.2732988} to create suggestions during workflow design.

\section{Conclusion}
\label{sec:conclusion}

In this paper we outlined the most basic principles of workflow provenance abstraction and  test-ran two abstraction systems over a real-world scientific workflow. Our analysis has shown that system-generated  and manual abstractions largely overlap from a process perspective, meanwhile, there is a dramatic mismatch from the data-respective. This has shown that automated abstraction  systems are skewed in their focus towards process, overlooking the data aspect. As a result they may find limited use in certain use-cases such as experiment reporting. Moreover, we observed that integrity policies of an abstraction system should be determined in light of end-purpose of abstractions.

\bibliographystyle{abbrv}
\bibliography{bib-cr}

\begin{thebibliography}{10}

\bibitem{summaryBigData13}
P.~Alper, K.~Belhajjame, C.~Goble, and P.~Karagoz.
\newblock Small is beautiful: Summarizing scientific workflows using semantic
  annotations.
\newblock In {\em Proceedings of the IEEE 2nd International Congress on Big
  Data (BigData 2013)}, Santa Clara, CA, USA, June 2013.

\bibitem{belhajjame-mina}
K.~Belhajjame, J.~Zhao, D.~Garijo, M.~Gamble, K.~Hettne, R.~Palma, E.~Mina,
  O.~Corcho, J.~M. Gomez-Perez, S.~Bechhofer, G.~Klyne, and C.~Goble.
\newblock {Using a suite of ontologies for preserving workflow-centric research
  objects}.
\newblock {\em Web Semantics: Science, Services and Agents on the World Wide
  Web}, 32(0):16 -- 42, 2015.

\bibitem{Biton2008}
O.~Biton, S.~Cohen-Boulakia, S.~B. Davidson, and C.~S. Hara.
\newblock {Querying and Managing Provenance through User Views in Scientific
  Workflows}.
\newblock In {\em Proceedings of the 24th International Conference on Data
  Engineering (ICDE)}, pages 1072--1081, 2008.

\bibitem{Cadenhead:2011:TPU:1998441.1998456}
T.~Cadenhead, V.~Khadilkar, M.~Kantarcioglu, and B.~Thuraisingham.
\newblock Transforming provenance using redaction.
\newblock In {\em Proceedings of the 16th ACM symposium on Access control
  models and technologies}, SACMAT '11, pages 93--102, New York, NY, USA, 2011.
  ACM.

\bibitem{Chapman:2008:EPS:1376616.1376715}
A.~P. Chapman, H.~V. Jagadish, and P.~Ramanan.
\newblock Efficient provenance storage.
\newblock In {\em Proceedings of the 2008 ACM SIGMOD International Conference
  on Management of Data}, SIGMOD '08, pages 993--1006, New York, NY, USA, 2008.
  ACM.

\bibitem{perera-cheney}
J.~Cheney and R.~Perera.
\newblock An analytical survey of provenance sanitization.
\newblock In {\em 5th International Provenance and Annotation Workshop,
  {IPAW}}, pages 113--126, Koln, Germany, 2014.

\bibitem{cohen-distill}
S.~Cohen-Boulakia, J.~Chen, P.~Missier, C.~Goble, A.~Williams, and
  C.~Froidevaux.
\newblock Distilling structure in taverna scientific workflows: a refactoring
  approach.
\newblock {\em BMC Bioinformatics}, 15(Suppl 1):S12, 2014.

\bibitem{myExp2008}
D.~De~Roure, C.~Goble, and R.~Stevens.
\newblock The design and realisation of the myexperiment virtual research
  environment for social sharing of workflows.
\newblock {\em Future Generation Computer Systems}, 25:561--567, May 2008.

\bibitem{Dey:2011:PTD:2032397.2032414}
S.~C. Dey, D.~Zinn, and B.~Lud\"{a}scher.
\newblock Propub: towards a declarative approach for publishing customized,
  policy-aware provenance.
\newblock In {\em {Proceedings of the 23rd international conference on
  Scientific and Statistical Database Management (SSDBM)}}, pages 225--243,
  Berlin, Heidelberg, 2011. Springer-Verlag.

\bibitem{DBLP:journals/fgcs/GarijoABCGG14}
D.~Garijo, P.~Alper, K.~Belhajjame, {\'{O}}.~Corcho, Y.~Gil, and C.~A. Goble.
\newblock Common motifs in scientific workflows: An empirical analysis.
\newblock {\em Future Generation Computer Systems}, 36:338--351, 2014.

\bibitem{HettneWBGMDRVGR12}
K.~M. Hettne, K.~Wolstencroft, K.~Belhajjame, C.~A. Goble, E.~Mina, et~al.
\newblock {Best Practices for Workflow Design: How to Prevent Workflow Decay}.
\newblock In {\em Proceedings of the 5th International Workshop on Semantic Web
  Applications and Tools for Life Sciences}, Paris, France, November 2012.

\bibitem{bpm-layers}
H.~Reijers and J.~Mendling.
\newblock {Modularity in Process Models: Review and Effects}.
\newblock In {\em Business Process Management}, volume 5240 of {\em Lecture
  Notes in Computer Science}, pages 20--35. Springer Berlin Heidelberg, 2008.

\bibitem{Starlinger:2014:SSS:2732977.2732988}
J.~Starlinger, B.~Brancotte, S.~Cohen-Boulakia, and U.~Leser.
\newblock Similarity search for scientific workflows.
\newblock {\em Proceedings of the VLDB Endowment}, 7(12):1143--1154, Aug. 2014.

\bibitem{sadi-semantic}
D.~Withers, E.~Kawas, L.~McCarthy, B.~Vandervalk, and M.~Wilkinson.
\newblock {Semantically-Guided Workflow Construction in Taverna: The SADI and
  BioMoby Plug-Ins}.
\newblock In {\em Leveraging Applications of Formal Methods, Verification, and
  Validation}, volume 6415 of {\em Lecture Notes in Computer Science}, pages
  301--312. Springer Berlin Heidelberg, 2010.

\end{thebibliography}

\appendix
\section{Appendix}
Figures displaying semi-automatically generated abstractions (black lines) overlayed on design abstractions (Taverna workflow screenshots).

\begin{figure}
\begin{center}
\includegraphics[scale=0.45]{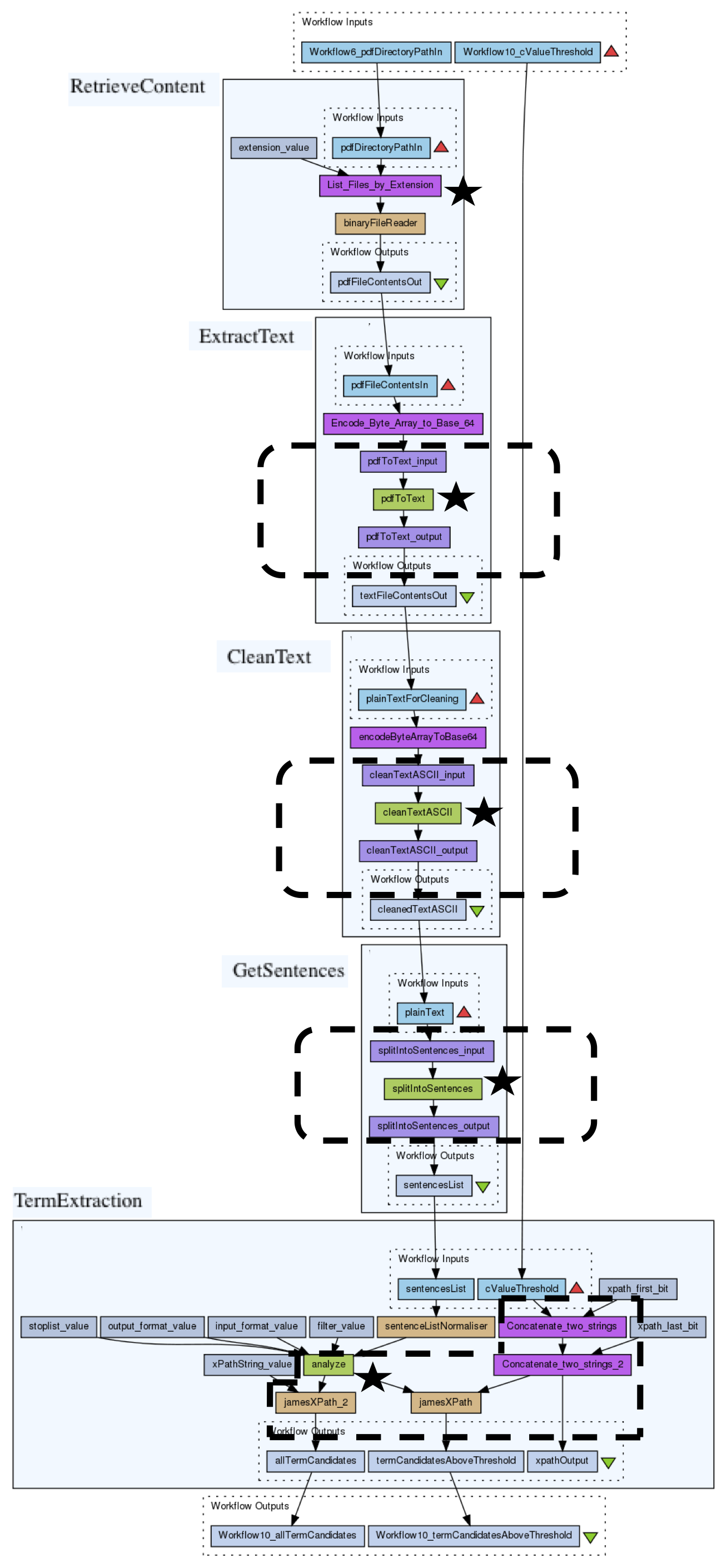}
\end{center}
\caption{Design Abstractions (sub-workflows) versus Workflow Summaries abstraction of selected adapters.}
\label{fig:collapse-w-example}
\end{figure}

\begin{figure}
\begin{center}
\includegraphics[scale=0.45]{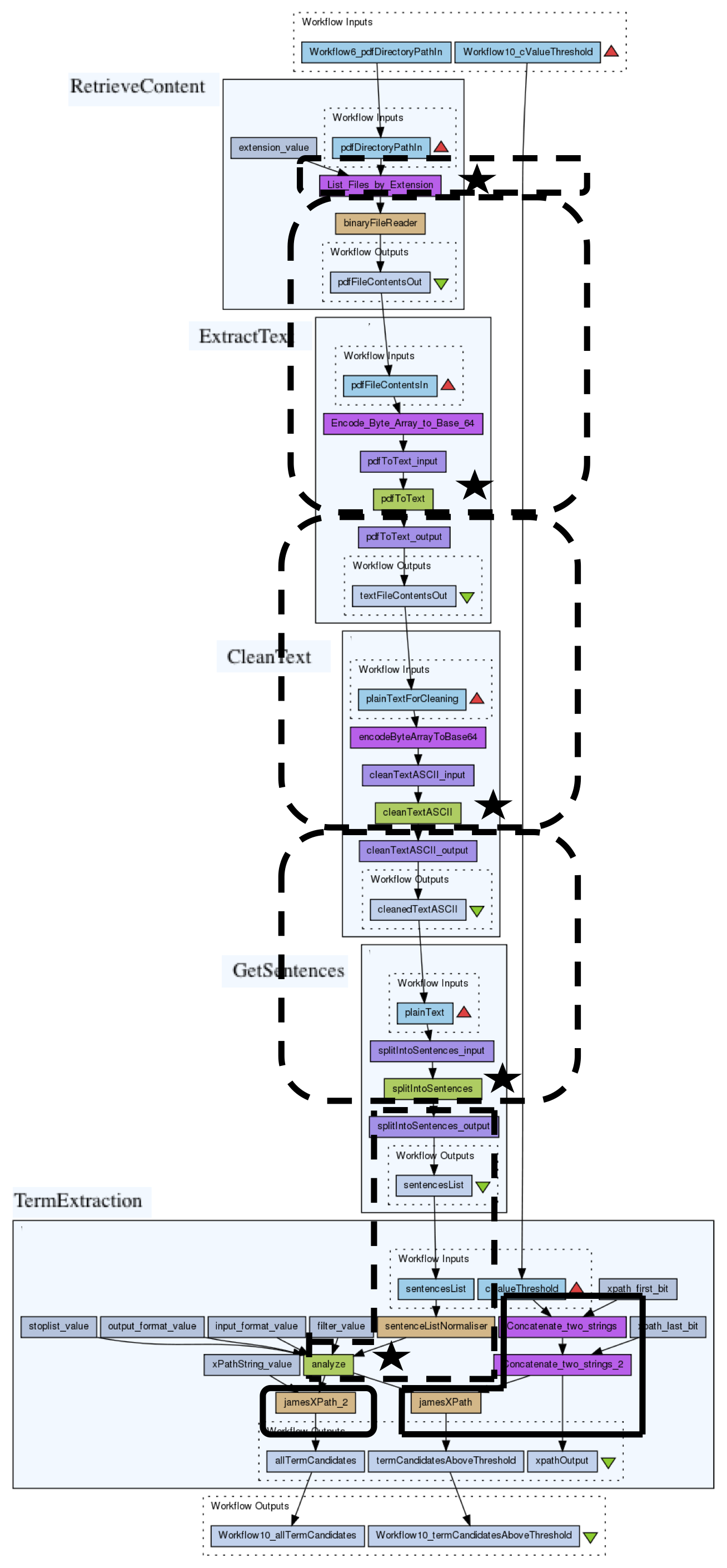}
\end{center}
\caption{Design Abstractions (sub-workflows) versus ZOOM  abstraction.}
\label{fig:zoom-example}
\end{figure}

\begin{figure}
\begin{center}
\includegraphics[scale=0.45]{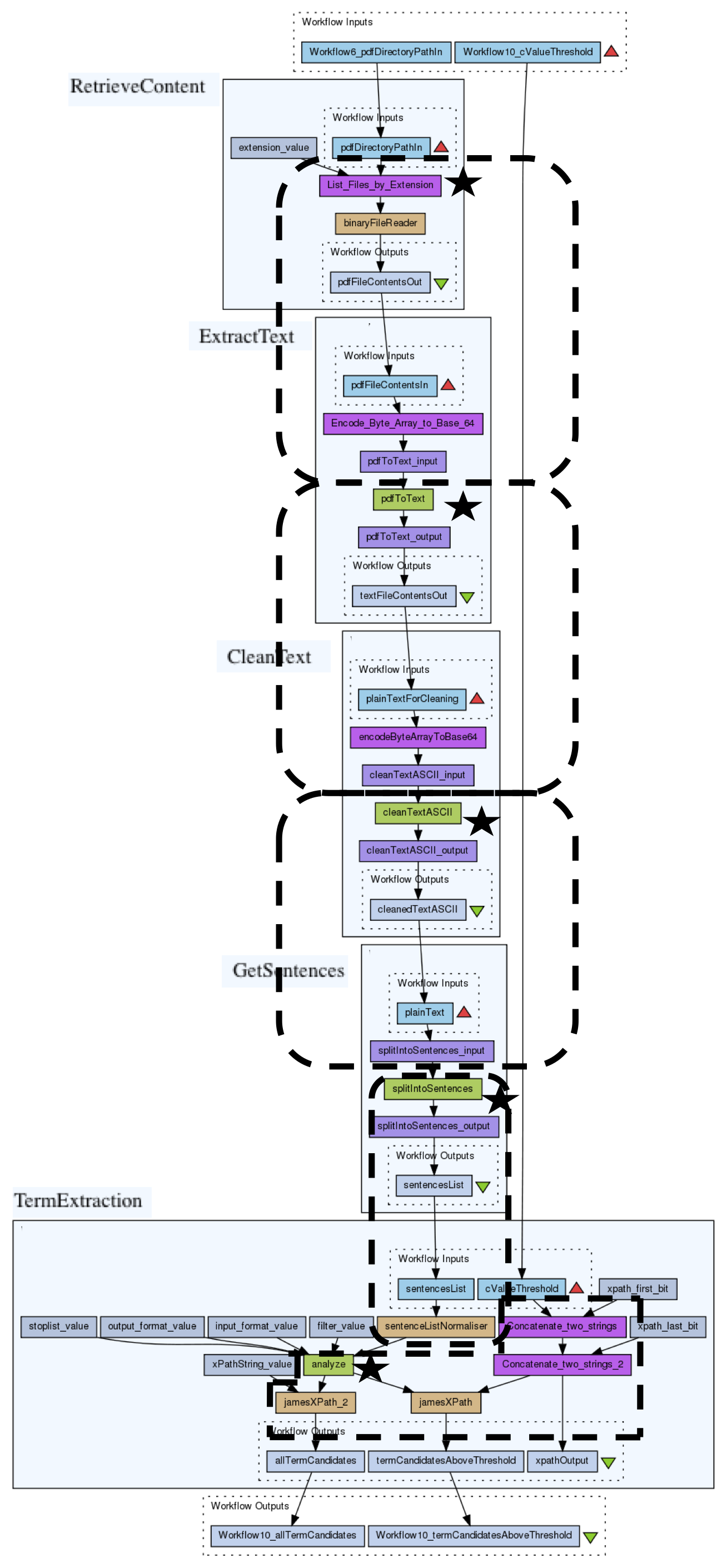}
\end{center}
\caption{Design Abstractions (sub-workflows) versus Workflow Summaries abstraction of all adapters.}
\label{fig:collapse-n-example}
\end{figure}

\end{document}